\documentclass[10pt]{article}
\usepackage[utf8]{inputenc}
\usepackage{amsmath}
\usepackage{graphicx,tabularx}
\usepackage{amsfonts,amsthm,eucal,amssymb,amscd}
\usepackage{xcolor}
\usepackage{epstopdf}
\usepackage{upgreek}
\usepackage{bbold} 
\usepackage{glossaries}

\definecolor{darkgray}{gray}{0.7}
\definecolor{lightgray}{gray}{0.3}
\definecolor{gray}{gray}{0.5}

\usepackage{microtype}

\begin{document}

\newtheorem{theo}{Theorem}[section]
\newtheorem{definition}[theo]{Definition}
\newtheorem{lem}[theo]{Lemma}
\newtheorem{prop}[theo]{Proposition}
\newtheorem{coro}[theo]{Corollary}
\newtheorem{exam}[theo]{Example}
\newtheorem{rema}[theo]{Remark}
\newtheorem{example}[theo]{Example}
\newcommand{\ninv}{\mathord{\sim}} 
\newtheorem{axiom}[theo]{Axiom}

\title{On the assumptions underlying KS-like contradictions}

\author{J. Acacio de Barros$^{1}$, Juan Pablo Jorge$^{2}$, and Federico Holik$^{3}$}

\date{\today}

\maketitle

\begin{center}

\begin{small}
1- School of Humanities and Liberal Studies, San Francisco State University, San Francisco, CA 94132, USA; barros@sfsu.edu\\
2- Physics Department, University of Buenos Aires, CABA (1428), Argentina; jorgejpablo@gmail.com\\
3- Instituto de F\'{i}sica La Plata, La Plata (1900),
Buenos Aires, Argentina; olentiev2@gmail.com\\
\end{small}
\end{center}

\vspace{1cm}

\begin{abstract}
\noindent The Kochen-Specker theorem is one of the fundamental no-go theorems in quantum theory. It has far-reaching consequences for all attempts trying to give an interpretation of the quantum formalism. In this work, we examine the hypotheses that, at the ontological level, lead to the Kochen-Specker contradiction. We emphasize the role of the assumptions about identity and distinguishability of quantum objects in the argument.
\end{abstract}
\bigskip
\noindent

\begin{small}
\centerline{\em Key words: Kochen-Specker theorem; quantum states;
quantum indistinguishability}
\end{small}

\section{Introduction}\label{s:Introduction}

One of the goals of physics is to predict a physical system's future behavior. For example, in classical mechanics, the current state of a particle allows for predicting its trajectory. Since points in a phase space completely determine classical states, it follows that all we need to know about a physical system is encoded in that space. Therefore, a subset of this space can represent any classical system's property: having a specific value of energy can be represented by the set of points in phase space that make the property actual. Suppose there is an epistemic uncertainty about a state. In that case, this uncertainty can be represented by probability measures over some particular subsets of phase space, namely, measurable subsets, usually taken to be the Borel sets. Thus, once the state of a system is known, its membership in the Borel set determines a proposition's truth-value. If a probabilistic description is given, the state determines the likelihood of any potential property.  

Things are quite different for quantum systems. A quantum system's pure state is not a point in phase space but a ray in a Hilbert space. This means that properties should be associated with rays or, more generally, with linear combinations of them, generating subspaces in the Hilbert space. This approach --and the study of the differences with classical mechanics-- was systematically studied by Birkhoff and von Neumann in \cite{BvN}. They found that its Hilbert space's closed subspaces can naturally represent a quantum system's elementary propositions. The main difference between quantum and classical propositions is that, while the latter form a distributive lattice, the former does not. This algebraic structure was called \textit{quantum logic}, due to its analogies to propositional calculus. Thus, while states of a classical probabilistic model can be represented as probability measures on the Boolean algebra of subsets of a phase space, the states of quantum systems can be described as measures over the quantum logic formed by the closed subspaces of its associated Hilbert space $\mathcal{H}$. Let us call this non-Boolean lattice $\mathcal{L}(\mathcal{H})$. Mathematically, $\mathcal{L}(\mathcal{H})$ is formed by all possible closed subspaces of the Hilbert space (equivalently, by its orthogonal projections). Conceptually, each element of $\mathcal{L}(\mathcal{H})$ represents an elementary property of the quantum system. Each of these elementary properties gives place to a YES-NO experiment, that tests whether the system has the property or not. These are called \emph{elementary tests}. As an example, consider the property ``the system has spin up in direction $\hat{z}$". An elementary test, is an experiment that gives the answer ``YES" if the system is detected to have spin up, and the answer ``NO" otherwise. 

The peculiar non-Boolean mathematical structure associated with the probabilistic quantum description of nature has its counterpart in the non-commutative algebraic character of quantum observables, represented by self-adjoint operators acting on a separable Hilbert space. By appealing to the spectral theorem, von Neumann successfully connected self-adjoint operators with the closed subspaces representing yes-no experiments. Indeed, for each maximal set $\mathcal{S}$ of compatible quantum observables defining a concrete measurement context, there exists a maximal Boolean subalgebra $\Sigma_{\mathcal{S}}$ of $\mathcal{L}(\mathcal{H})$, representing the events of a classical probability space associated to that context. Any quantum state defines a classical probability space when combined with $\Sigma_{\mathcal{S}}$. Notice that an important way of mathematically describing the quantum observables is by appealing to the notion of \textit{partial Abelian algebra}. Intuitively, this means that observables bear a compatibility relation among them: some of them commute (when they are compatible), but in general, they will not commute.

The Kochen-Specker (KS) theorem \cite{KST} exploits the fact that quantum observables form a partial Abelian algebra and is one of the fundamental no-go theorems in quantum theory. The KS theorem was conceived initially to discard specific families of hidden-variable models. Its consequences were far-reaching in the literature on the interpretation of the quantum formalism (see, for example, the discussion in \cite{Newton-Olimpia}). It can be extended to many probabilistic models of interest \cite{Doring-KSVNA,Svozil-KS,Smith-KS}, so its applicability domain goes far beyond standard quantum mechanics. The study of contextual systems outside the quantum domain \cite{AcacioContextuality-2015,Aerts,Khrennikiv-Ubiquitous,Holik-Probabilities-Generalized} can be related to the interpretation of the KS result.

The KS theorem imposes severe restrictions on the possible valuations that can be defined over the propositions associated with quantum systems. In this work, we discuss the assumptions that underlie the KS contradiction at the ontological level. In particular, we focus on the role played by the identity of quantum objects. As is well known, quantum indistinguishability \cite{Holik-2020-Non-individuals} poses a severe threat to the assumption that quantum systems are individuals in the traditional sense (as billiard balls are). The peculiar behavior of quantum systems regarding the impossibility of labeling or identifying them leads many authors to conclude that quantum systems of the same kind are utterly indistinguishable and that they cannot be considered as individuals at all. This leads to the idea of developing an ontology based on \textit{non-individuals}, which are genuinely indistinguishable objects that cannot be identified, and for which the standard theory of identity does not apply (see for example, \cite{Schro-Non-individuals,French-Krause-2006}). In this work, we discuss the implications of the assumption that quantum systems are non-individuals for deriving the Kochen-Specker contradiction (abbreviated as ``KS contradiction", in what follows). See also \cite{deBarros-Holik-Krause-2017,DeBarros-Indistinguishability-2019,deBarros-IndistinguishabilityandNegative} for more discussion on this. 

The paper is organized as follows. We start by reviewing some elementary facts about probabilities in quantum and classical theories in Section \ref{s:ElementaryFacts}. In section \ref{KSTheorem}, we review the KS contradiction. Then, in Section \ref{s:Indistinguishability}, we present a discussion about quantum particles' indistinguishability and establish the main features of an ontology based on non-individuals. In Section \ref{s:IdentityInKS}, we show how the assumptions about quantum systems' identity are used in the KS argument. Finally, in Section \ref{s:Conclusions}, we draw our conclusions.

\section{Quantum States and the lattice of propositions of physical systems}\label{s:ElementaryFacts}

In this section, we give some elementary technical definitions needed to understand the Kochen-Specker contradiction.

\subsection{Classical probabilities}\label{s:ClassicalProba}

It will be important for us to first recall how probabilities are defined
in a classical setting. Given a set $\Omega$, let us consider
a $\sigma$-algebra $\Sigma\subseteq\mathcal{P}(\Omega)$ of subsets of it.
Then, a \textit{probability measure} will be given by a
function
\begin{equation}\label{e:kolmogorovian}
\mu:\Sigma\rightarrow[0,1]
\end{equation}
satisfying the following axioms. 
\begin{enumerate}

\item $\mu(\emptyset)=0$


\item for any
pairwise disjoint and denumerable family
$\{A_{i}\}_{i\in\mathbb{N}}$,
$\mu(\bigcup_{i}A_{i})=\sum_{i}\mu(A_{i})$.

\end{enumerate}

A $\mu$ in \eqref{e:kolmogorovian} satisfying axioms 1 and 2 is very useful since it captures the general features of probabilities in many examples of interest. Probabilities defined by \eqref{e:kolmogorovian} are usually called Kolmogorovian \cite{KolmogorovProbability}. In order to fix ideas, let us consider the concrete example of a dice. Each possible outcome can be represented by an element of the set $\Omega=\{1,2,3,4,5,6\}$. Let $\mathcal{P}(\Omega)$ represent all possible subsets of $\Omega$. Before throwing the dice, consider the event ``the outcome is even.'' It is naturally represented by the subset $\{2,4,6\}$. Alternatively, ``the outcome is greater than $3$'' is represented by $\{4,5,6\}$. Similarly, any possible event can be represented as a subset of $\mathcal{P}(\Omega)$ (in this example, the $\sigma$-algebra $\Sigma$ is equal to $\mathcal{P}(\Omega)$). A probability measure $\mu$ assigns a real number to each member of $\mathcal{P}(\Omega)$, in such a way that the rules listed above are satisfied. The uniform probability is defined by the condition $\mu(\{i\})=\frac{1}{6}$, for $i=1,...,6$. Nevertheless, other measures can be considered (in order to represent, for example, loaded dices). One key feature of $\Sigma$ is that it is a Boolean algebra (as can be straightforwardly checked for $\mathcal{P}(\Omega)$ in the dice example). This means that, in $\Sigma$, we can form conjunctions (given by set-theoretical intersections), disjunctions (set-theoretical unions), and complements (set-theoretical complements), and that these logical operations satisfy some specific algebraic properties. In particular, the fact that $\Sigma$ is a Boolean algebra implies that there will always exist deterministic valuations: a function $\mu(P)\in\{0,1\}$, assigning truth values for all $P\in\mathcal{P}(\Omega)$ (and satisfying axioms \ref{e:kolmogorovian}). Geometrically speaking, all possible probabilistic states form a convex set with the uniform probability lying in its centroid, and deterministic assignments (i.e., classical truth value assignments) can be represented as its extreme points.

\subsection{Quantum probabilities and observables}\label{s:QPandObservables}

During the '30s, Birkhoff and von Neumann studied the propositional structures associated with quantum systems and compared them to classical systems. This study gave birth to a mathematical structure known as \textit{quantum logic} \cite{BvN}. According to the quantum logical approach, an elementary experiment associated with a quantum system is given by a yes-no test, i.e., a test in which we get the answer ``YES'' or the answer ``NO'' \cite{piron}, and it is mathematically represented by a closed subspace of the Hilbert space. Conceptually, each elementary test, does the job of testing whether the system has a certain property or not. Thus, to each family of equivalent elementary tests (i.e., experiments that test the same physical property), we can assign an elementary property of the system involved. 

In order to illustrate this, let us consider an example. Suppose that a quantum system has a Hamiltonian $H$, and we are interested in testing whether the system has energy value $\epsilon$ or not. If the system is prepared in an eigenstate $|\psi\rangle$, i.e., $H|\psi\rangle=\epsilon|\psi\rangle$, then the probability of obtaining $\epsilon$ after measuring its energy is one. In that sense, the state $|\psi\rangle$ makes the proposition ``the system has energy $\epsilon$" true, and we say that the system possesses the property described by that proposition. Any other eigenvector of $H$ with eigenvalue $\epsilon$ does the same, as well as any linear combination of them. Thus, we see that the closed subspace $\mathbb{S}_{\epsilon}$ -- spanned by all eigenstates with eigenvalue $\epsilon$ -- can be used to represent the property ``the system has energy $\epsilon$.'' Equivalently, one can use the projection operator $P_{\epsilon}$ associated to $\mathbb{S}_{\epsilon}$, given the fact that closed spaces and orthogonal projections are in a one to one correspondence. In the following, remember that an operator $P\in\mathcal{B}(\mathcal{H})$ is said to be an orthogonal projection if it is self-adjoint and satisfies $P^{2}=P$.

Denote by $\mathcal{B}(\mathcal{H})$ the set of bounded operators acting on $\mathcal{H}$ and let $\mathcal{P}(\mathcal{H})$ be the set of all orthogonal projections acting on $\mathcal{H}$. It is possible to show that $\mathcal{P}(\mathcal{H})$ can be endowed with an orthocomplemented lattice structure $\mathcal{L}(\mathcal{H})= \langle\mathcal{P}(\mathcal{H}),\ \wedge,\ \vee,\ \neg,\ \mathbf{0},\ \mathbf{1}\rangle$, where $P\wedge Q$ is the orthogonal projection associated with the intersection of the ranges of $P$ and $Q$, $P\vee Q$ is the orthogonal projection associated with the closure of the direct sum of the ranges of $P$ and $Q$, $\mathbf{0}$ is the null operator (the bottom element of the lattice), $\mathbf{1}$ is the identity operator (the top element), and $\neg(P)$ is the orthogonal projection associated with the orthogonal complement of the range of $P$ \cite{mikloredeilibro}. Alike the Boolean algebras used in classical probabilistic models, $\mathcal{P}(\mathcal{H})$ is a non-Boolean lattice. It is always modular in the finite-dimensional case and never modular in the infinite one \cite{mikloredeilibro}.  

In quantum mechanics, states can be considered as functions that
assign probabilities to the elements of $\mathcal{L}(\mathcal{H})$.
A state on a quantum system is represented by a function \cite{Redei-Summers2006,Hamhalter}:
\begin{equation}\label{e:nonkolmogorov}
\mu:\mathcal{L}(\mathcal{H})\longrightarrow [0,1]
\end{equation}
satisfying:
\begin{enumerate}

\item $\mu(\textbf{0})=0$.


\item For any pairwise orthogonal and denumerable family $\{P_{j}\}_{j\in\mathbb{N}}$,
$\mu(\bigvee_{j}P_{j})=\sum_{j}\mu(P_{j})$.
\end{enumerate}

Gleason's theorem \cite{Gleason,Gleason-Dvurechenski-2009} implies that whenever $dim(\mathcal{H})\geq 3$, the convex set $\mathcal{C}(\mathcal{L}(\mathcal{H}))$ of all measures of the form \eqref{e:nonkolmogorov} can be put in a one to one correspondence with the set $\mathcal{S}(\mathcal{H})$ of all
positive and trace-class operators of trace one acting in $\mathcal{H}$. The correspondence is such that for every measure $\mu$ satisfying the above axioms, there exists a density operator
$\rho_{\mu}\in\mathcal{S}(\mathcal{H})$ such that for every
orthogonal projector $P$ representing an elementary test, we have
\begin{equation}\label{e:bornrule}
\mu(P)=\mbox{tr}(\rho_{\mu}P).
\end{equation}

In quantum mechanics, any observable quantity can be represented by a self-adjoint operator. For every self-adjoint operator $A$, if the system is prepared in the state $\rho$, its mean value is given by the formula
\begin{equation}
\langle A\rangle=\mbox{tr}(\rho A).
\end{equation}

Due to the spectral theorem, the set of all self-adjoint operators can be put in one--one correspondence with projective valued measures (PVM) \cite{vN}. Let $\mathcal{B}(\mathbb{R})$ be the Borel set of the real line. Given a self-adjoint operator $A$, its {projection-valued measure} $M_{A}$
is a map \cite{mikloredeilibro}
\begin{equation}
M_{A}: \mathcal{B}(\mathbb{R})\mapsto \mathcal{P}(\mathcal{H}),
\end{equation}
such that
\begin{enumerate}

\item $M_{A}(\emptyset)=\mathbf{0}$

\item $M_{A}(\mathbb{R})=\mathbf{1}$

\item $M_{A}(\cup_{j}(B_{j}))=\sum_{j}M_{A}(B_{j})$,\,\, \mbox{for any mutually disjoint
family} ${B_{j}}$

\item $M_{A}(B^{c})=\mathbf{1}-M_{A}(B)=(M_{A}(B))^{\bot}$.

\end{enumerate}
Propositions about the observable $A$ are naturally represented by Borel subsets of $\mathbb{R}$. Given $\Delta\in\mathcal{B}(\mathbb{R})$, w can always associate a unique proposition ``the value of $A$ lies in $\Delta$.'' Thus, $M_{A}$ connects the propositions about $A$ with the properties of the quantum system, represented Hilbert space's closed subspaces (or, equivalently, with the orthogonal projections). Several significant consequences follow from this association. 

First, it is straightforward to show that the range of $M_{A}$ is a Boolean subalgebra $\Sigma_{A}$ of $\mathcal{L}(\mathcal{H})$. This implies that each observable defines a classical propositional system, in a similar way as we described in the example of the dice presented in Section \ref{s:ClassicalProba}. When we consider a maximal observable (specified by a measurement context), the corresponding Boolean subalgebra is maximal \cite{piron}. For example, a non-degenerate observable of a six-dimensional quantum system defines an outcome space formally equivalent to that of a dice. These considerations imply that quantum observables define classical random variables when considered in connection to their respective measurement contexts. Notice that a quantum state defines a classical probability for each measurement context. In terms of the ``dice example", each measurement context behaves as a loaded dice, with the probabilities fioxed by the global quantum state.

Second, the collection of all possible propositions --represented by $\mathcal{L}(\mathcal{H})$-- can be described as a pasted family of (maximal) Boolean algebras. All possible properties associated with a given quantum system can be grouped in different Boolean algebras, representing the different measurement contexts. Two properties are compatible if and only if they can be included in a common Boolean algebra (i.e., tested in the same measurement context). This compatibility connection gives place to the mathematical notion of \textit{partial Boolean algebra}. Furthermore, the different Boolean algebras are such that they share things in common. For example, in a two-parties scenario, the observable $A\otimes B$ will not commute with $A\otimes C$, whenever $[B,C]\neq 0$. However, the observable $A\otimes I$ commutes with $A\otimes B$ and $A\otimes C$. Thus, we find the same observable in different measurement contexts. This means that quantum contexts define \textit{intertwined} Boolean algebras. Moreover, in general, the intertwining will be very complex (and difficult to characterize). This means that $\mathcal{L}(\mathcal{H})$, the collection of all possible elementary quantum properties, can be considered as a \textit{pasting} of intertwined Boolean algebras. As we will see in the following section, the fact that the different measurement contexts (or maximal Boolean subalgebras) share common elements lies at the core of the KS contradiction.


\section{Kochen-Specker theorem}\label{KSTheorem}

In this section, we review the KS theorem \cite{KST,Svozil-KS,Smith-KS}. It depends critically on the non-Boolean character of $\mathcal{L}(\mathcal{H})$ described in the previous section. Kochen and Specker aimed to study the possibility of finding hidden variables for quantum mechanics. They focused on a very particular family of hidden-variable models. Namely, they took as a model the relationship between classical statistical mechanics and thermodynamics. As a result of their research, it turns out that no hidden-variable theory of this sort can exist for the quantum case. 

In classical statistical mechanics, there is a hidden (i.e., non-observable) state space $\Omega$ possessing micro-states $\lambda\in\Omega$, and there exists a probability distribution $p(\lambda)$ (that defines the probabilistic state of the system). For each macroscopic observable $A$, a random variable $f_{A}:\Omega\longrightarrow\mathbb{R}$ is assigned in such a way that $\langle A\rangle=\int_{\Omega}f_{A}(\lambda)p(\lambda)d\lambda$. In other words, the values of macroscopic observables can be computed as mean values of random variables defined over $\Omega$ using the usual probabilistic formulas. Notice that each $\lambda\in \Omega$ assigns a value $v_{A}$ to each observable $A$ according to the formula $v_{\lambda}(A)=f_{A}(\lambda)$. And this assignment is such that a functional condition is satisfied in the following sense: the value assigned to a function of an observable is given by the function evaluated in the value of the given observable. In formulae, this condition reads
\begin{equation}\label{e:TruthFunctionalityOperators}
B=G(A)\,\,\Longrightarrow\,\,v_{\lambda}(B)=g(v_{\lambda}(A))  
\end{equation}

\noindent for all $\lambda\in\Omega$, where $G:\mathcal{A}\longrightarrow\mathcal{A}$ is a Borel function that maps observables in observables, and by $v_{\lambda}(B)=g(v_{\lambda}(A))$, we mean a map $g:\mathbb{R}\longrightarrow\mathbb{R}$ with the same functional form as $G$. 
As an example, if $B=A^{2}$, then $v_{\lambda}(B)=v_{\lambda}(A)^{2}$ (if $v_{\lambda}(A)=2$, then
$v_{\lambda}(B)=4$). This condition is usually called FUNC in the literature (see for example \cite{Isham}), and it expresses the fact that observables are not all independent, and neither are the values assigned to them. Hidden variables satisfying the FUNC condition are the reasonable candidates for Kochen and Specker. Thus, they look for hidden variables satisfying the FUNC condition, and such that
\begin{equation}\label{e:MeanHidden}
\langle A\rangle:=\mbox{tr}(\hat{\rho}\hat{A})=
\int_{\Omega}f_{A}(\lambda)p_{\rho}(\lambda)d\lambda
\end{equation}
for every quantum observable $A$ and every quantum state $\rho$. Notice that each quantum state $\rho$ has its counterpart in the classical probability distribution $p_{\rho}$.

Let us now study with more detail how the hidden variables --assumed to exist-- should assign values to the observables. Let $\mathcal{A}(\mathcal{H})$ be the set of all self-adjoint operators acting on the Hilbert space $\mathcal{H}$. Any $\lambda\in\Omega$ defines a valuation function $v_{\lambda}:\mathcal{A}(\mathcal{H})\longrightarrow\mathbb{R}$, by appealing to the assignment $v_{\lambda}(A)=f_{A}(\lambda)$. A function such as $v_{\lambda}$ can be called a \textit{prediction
function} (see section II in \cite{KST}), because it assigns a given
value to each quantum observable. 

Let us see how this works for elementary properties represented by projection operators (see also sections I and II in \cite{KST}). As is well known, two quantum
mechanical observables represented by self-adjoint operators $A$ and
$B$, respectively, are compatible, if and only if, there exist Borel
functions $g_{1}$ and $g_{2}$, and a self adjoint operator $C$, such
that $A=g_{1}(C)$ and $B=g_{2}(C)$. Thus, whenever $A$ and $B$ are
compatible, using the functionality condition
\eqref{e:TruthFunctionalityOperators} we have
$v_{\lambda}(AB)=v_{\lambda}(g_{1}(C)g_{2}(C))=v_{\lambda}((g_{1}g_{2})(C))=(g_{1}g_{2})(v_{\lambda}(C))=g_{1}(v_{\lambda}(C))g_{2}(v_{\lambda}(C))=v_{\lambda}(g_{1}(C))v_{\lambda}(g_{2}(C))=v_{\lambda}(A)v_{\lambda}(B)$. If $\alpha$ and $\beta$ are real numbers, we also have, for
compatible $A$ and $B$, that $v_{\lambda}(\alpha A+\beta B)=v_{\lambda}(\alpha
g_{1}(C)+\beta g_{2}(C))=v_{\lambda}((\alpha g_{1}+\beta g_{2})(C))=(\alpha
g_{1}+\beta g_{2})(v_{\lambda}(C))=\alpha g_{1}(v_{\lambda}(C))+\beta g_{2}(v_{\lambda}(C))=\alpha
v_{\lambda}(g_{1}(C))+\beta v_{\lambda}(g_{2}(C))=\alpha v_{\lambda}(A)+\beta v_{\lambda}(B)$. It
follows that, for each $\lambda\in \Omega$, $v_{\lambda}$ defines a partial Abelian algebra homomorphism.
Furthermore, if $P^{2}=P$ and $P^{\dag}=P$ (i.e., if $P$ is an
orthogonal projection), we have
$v_{\lambda}(P)=v_{\lambda}(P^{2})=v_{\lambda}(P)v_{\lambda}(P)=v_{\lambda}(P)^{2}$, and then, $f_{P}=0$ or
$f_{P}=1$. In words: each hidden state assigns an homomorphism $v_{\lambda}:\mathcal{L}(\mathcal{H})\longrightarrow \{0,1\}$. This means that $v_{\lambda}$ assigns truth values ($0$ or $1$) to each proposition in $\mathcal{L}(\mathcal{H})$ in a functional way (see \cite{Jorge-Holik-2020} for the technical meaning of ``functional"). Let us illustrate how this last condition (and \ref{e:TruthFunctionalityOperators}) work in a given measurement context. Every measurement context can be represented by a maximal collection $\{P_{i}\}$ of mutually orthogonal (i.e., $P_{i}P_{j}=\mathbf{0}$, whenever $i\neq j$) one dimensional projection operators, that form a resolution of the identity 

\begin{equation}
\sum_{i}P_{i}=\mathbf{1}    
\end{equation}

\noindent Then, it is easy to check that

\begin{equation}\label{e:TruthValueResolutionIdentity}
\sum_{i}v_{\lambda}(P_{i})=1    
\end{equation}

\noindent Notice that the above equation, implies that, if in a given measurement context $v_{\lambda}(P_{i})=1$, then,  $v_{\lambda}(P_{j})=0$, for $j\neq i$). Thus, we arrive at the conclusion that, if hidden variables satisfying FUNC and \ref{e:MeanHidden} exist, there should also exist valuations 
$v:\mathcal{L}(\mathcal{H})\longrightarrow\{0,1\}$ having the
property that $\sum_{i}v(P_{i})=1$ for any family
$\{P_{i}\}_{i\in\mathbb{N}}$ of one dimensional orthogonal elements
of $\mathcal{L}(\mathcal{H})$ satisfying $\sum_{i}P_{i}=\mathbf{1}$. But Kochen and Specker show that this is impossible. In order to see why, let us consider a simple example (presented in \cite{Cabello-SimpleKS}) of why such valuations cannot exist.

Consider a four dimensional quantum model and the following nine measurement contexts:

\begin{align}\label{e:KSCabello}
\textcolor{blue}{\hat{P}_{0,0,0,1}}+\textcolor{darkgray}{\hat{P}_{0,0,1,0}}+\textcolor{red}{\hat{P}_{1,1,0,0}}+\textcolor{lightgray}{\hat{P}_{1,-1,0,0}}
& =\hat{1}\\\nonumber
\textcolor{blue}{\hat{P}_{0,0,0,1}}+\textcolor{yellow}{\hat{P}_{0,1,0,0}}+\hat{P}_{1,0,1,0}+\textcolor{green}{\hat{P}_{1,0,-1,0}}
& =\hat{1}\\\nonumber
\textcolor{cyan}{\hat{P}_{1,-1,1,-1}}+\textcolor{gray}{\hat{P}_{1,-1,-1,1}}+\textcolor{red}{\hat{P}_{1,1,0,0}}+\textcolor{teal}{\hat{P}_{0,0,1,1}}
& =\hat{1}\\\nonumber
\textcolor{cyan}{\hat{P}_{1,-1,1,-1}}+\textcolor{magenta}{\hat{P}_{1,1,1,1}}+\textcolor{green}{\hat{P}_{1,0,-1,0}}+\textcolor{olive}{\hat{P}_{0,1,0,-1}}
& =\hat{1}\\\nonumber
\textcolor{darkgray}{\hat{P}_{0,0,1,0}}+\textcolor{yellow}{\hat{P}_{0,1,0,0}}+\textcolor{purple}{\hat{P}_{1,0,0,1}}+\textcolor{brown}{\hat{P}_{1,0,0,-1}}
& =\hat{1}\\\nonumber
\textcolor{gray}{\hat{P}_{1,-1,-1,1}}+\textcolor{magenta}{\hat{P}_{1,1,1,1}}+\textcolor{brown}{\hat{P}_{1,0,0,-1}}+\textcolor{violet}{\hat{P}_{0,1,-1,0}}
& =\hat{1}\\\nonumber
\textcolor{orange}{\hat{P}_{1,1,-1,1}}+\textcolor{pink}{\hat{P}_{1,1,1,-1}}+\textcolor{lightgray}{\hat{P}_{1,-1,0,0}}+\textcolor{teal}{\hat{P}_{0,0,1,1}}
& =\hat{1}\\\nonumber
\textcolor{orange}{\hat{P}_{1,1,-1,1}}+\textcolor{lime}{\hat{P}_{-1,1,1,1}}+\hat{P}_{1,0,1,0}+\textcolor{olive}{\hat{P}_{0,1,0,-1}}
& =\hat{1}\\\nonumber
\textcolor{pink}{\hat{P}_{1,1,1,-1}}+\textcolor{lime}{\hat{P}_{-1,1,1,1}}+\textcolor{purple}{\hat{P}_{1,0,0,1}}+\textcolor{violet}{\hat{P}_{0,1,-1,0}}
& =\hat{1},
\end{align}

Each equation above represents a different measurement context. For example, in the first line, we have a complete observable with four possible outcomes, represented by the projection operators $\hat{P}_{0,0,0,1}$, $\hat{P}_{0,0,1,0}$, $\hat{P}_{1,1,0,0}$ and $\hat{P}_{1,-1,0,0}$ (the subindexes represent the coordinates of rays in the Hilbert space). But the one-dimensional projections are chosen so that each one of them is repeated in different lines, something that is closely related to what we have discussed in section \ref{s:QPandObservables} about the intertwining of contexts. The repeated projections are painted with the same color in \ref{e:KSCabello}. Assuming the existence of a functional valuation to the set $\{0,1\}$ satisfying \ref{e:TruthValueResolutionIdentity}, we reach a contradiction (see \cite{Cabello-SimpleKS} for details). This is so because, by summing-up all the valuations of the equations in \ref{e:KSCabello}, since each projection is repeated twice, we obtain an even number in the left, but an odd number in the right, and that can never be an equality! The intertwining is responsible for the contradiction if the values assigned to each property are preserved among the different contexts. In other words, in order to reach the contradiction, we must assume that $\hat{P}_{0,0,0,1}$ in the first line of \ref{e:KSCabello} is the same as $\hat{P}_{0,0,0,1}$ when considered in the second line, and thus, that it retains its value assignment among the different contexts. A similar consideration holds for the remaining properties listed in \ref{e:KSCabello}.

Thus, we see that an implicit assumption behind the KS contradiction is that the quantum system can be \textit{identified} among the different contexts. Thus, its properties must retain the value assignments given by the hidden parameters $\lambda$. Is it legitimate to make these identifications? Can quantum systems be truly identified? We will elaborate on this remark in section \ref{s:IdentityInKS}, after discussing the problem of quantum indistinguishability in the next section.

\section{Quantum Indistinguishability}\label{s:Indistinguishability}

The symmetrization postulate \emph{is an independent postulate of standard QM} \cite{Holik-2020-Non-individuals,Omar-2005}. When we consider two quantum systems that are of the same kind, we must impose a symmetrization condition in their states. It turns out that, at the fundamental level, only two possibilities are experimentally observed. Either the systems are bosons, and their quantum state is symmetric under permutation, i.e.,
\begin{equation}
|\psi\rangle_{+}=\frac{1}{\sqrt{2}}(|a\rangle\otimes|b\rangle+|b\rangle\otimes|a\rangle)\leftarrow\mbox{(\emph{Bosons})},    
\end{equation}
or they are Fermions, and their wave function is antisymmetric, namely
\begin{equation}
|\psi\rangle_{-}=\frac{1}{\sqrt{2}}(|a\rangle\otimes|b\rangle-|b\rangle\otimes|a\rangle)\leftarrow\mbox{(\emph{Fermions})}.    
\end{equation}

The symmetrization postulate and its consequences have been empirically tested with great accuracy. It is a purely quantum feature --that has to be added to entanglement and superposition-- and it gives place to quantum statistics. As a result, the permutation of two quantum systems of the same kind gives no physical difference. In turn, this implies that quantum systems cannot be consistently identified (and re-identified in time) \cite{Holik-2020-Non-individuals}. These unusual physical features were quickly recognized as a problematic aspect concerning the assumption that quantum systems are individuals \cite{French-Krause-2006}. The Nobel laureate physicist Erwin Schr\"{o}dinger stated this point emphatically:

\begin{quotation}``I mean this: that the elementary particle is not an individual; it
cannot be identified, it lacks `sameness'. The fact is known to
every physicist, but is rarely given any prominence in surveys
readable by nonspecialists. In technical language it is covered by
saying that the particles `obey' a new fangled statistics, either
Einstein-Bose or Fermi-Dirac statistics. [...] The implication, far
from obvious, is that the unsuspected epithet `this' is not quite
properly applicable to, say, an electron, except with caution, in a
restricted sense, and sometimes not at all." (\cite{Schro-Non-individuals}, p.197)\end{quotation}

\noindent In this way, Shr\"{o}dinger points his finger towards an
\emph{alternative} ontology with regards to individuality:

\begin{quote}
``[We are] compelled to dismiss the idea that (\ldots) a particle is
an individual entity which retains its ``sameness" forever. Quite on
the contrary, we are now obliged to assert that the ultimate
constituents of matter have no ``sameness" at all.
I beg to emphasize this and I beg you to believe it: It is not a
question of our being able to ascertain the identity in some
instances and not being able to do so in others. It is beyond of
doubt that the question ``sameness" or ``identity", really and truly
has no meaning." (\cite{Schro-Non-individuals}, p.197)
\end{quote}

Indistinguishability is a distinctive physical feature of quantum mechanical systems, and we emphasize that it is an independent postulate of standard QM, different from entanglement. There can be disentangled indistinguishable systems, and there can be entangled systems that are perfectly distinguishable. Indistinguishability lies behind superconductivity and Bose-Einstein condensation. With the advent of quantum information theory, it was quickly recognized as a technological resource \cite{HOM-1987,Omar-2002,LoFranco-PRL-2018,Castellini-2019}. Thus, indistinguishability (or non-individuality) is a concept that leads to a positive physical principle. It allows physicists to conceive new physics, and it can be used to develop technological devices (e.g., as a resource in quantum information theory). Thus, why not take it seriously, at the ontological level? There are alternative interpretations (i.e., Bohmian mechanics) in which particles are explicit individual entities. However, it is also interesting to explore other possibilities that are close to common practice in physics and explore the assumption that there are, in nature, objects that are different \emph{solo numero} and that cannot be identified, as Schr\"{o}dinger suggested. Here, we explore the implications of non-individuality for the discussion about the KS contradiction.

\section{KS contradiction and identity}\label{s:IdentityInKS}

In this section, we discuss the role of quantum systems' identity in the derivation of the KS contradiction. Before entering the KS case, we will discuss the EPR example to motivate our point.

\subsection{Identity in the EPR argument}

In the famous EPR argument \cite{EPR-paper}, a bipartite system is assumed to be prepared in an entangled state. The authors adopt the following line of thought. Alice and Bob are two physicists working in labs that are far apart from each other. They each observe a particle arriving in their lab simultaneously (in a reference frame at rest with respect to Alice and Bob's labs). 

\begin{itemize}
\item \textbf{Step 1:} Alice is free to decide if she measures position or momentum in her system. Both measurements are mutually incompatible; they define different measurement contexts and cannot be performed simultaneously.

\item \textbf{Step 2:} If Alice decides to measure position, she can infer that Bob's system possesses a defined position. Once Alice's result is given, it is also possible to predict Bob's result with certainty. Then, according to the EPR reality criteria, Bob's system's position is an element of reality, given that the systems are very far away, and there cannot be any mechanical disturbance due to Alice's measurement. A similar consideration follows if Alice decides to measure momentum in her system.

\item\textbf{Step 3:} By \textit{identifying} the results of the two (mutually exclusive) possibilities of measurement, \emph{it follows} that both magnitudes must be elements of reality. Furthermore, since both refer to Bob's particle, they must be elements of reality \textit{simultaneously}.  
\end{itemize}

The authors of the EPR paper are very clear about the act of identification. They say that ``...it is possible to assign two different wave functions (in our example $\psi_{m}$ and $\phi_{r}$) to the same reality (the second system after the interaction with the first)" \cite{EPR-paper}. They refer to Bob's particle, when considered in two different measurement contexts, as ``the same reality". In this way, they assume that the quantum system can be identified among the different contexts. If this were true, quantum mechanics would be incomplete, as they correctly conclude. However, one may ask: are quantum systems identifiable? This problem is widely studied in the philosophy of logic. The notion of transmundane identity - ``identity between possible worlds" - is the notion that the same object exists in more than one possible world (with the real world treated as one of the possible worlds). Therefore, one has one's home in a framework of ``possible worlds'' to analyze, or at least paraphrase, statements about what is possible or necessary \cite{Kripke}.

The issue of transmundane identity has been highly controversial, even among philosophers who accept the legitimacy of speaking of possible worlds. Opinions range from viewing the notion of an identity held between objects in different possible worlds as so problematic, that it is unacceptable to viewing the notion as entirely innocuous, and no more problematic than the claim that objects individuals could have existed with somewhat different properties. Things are complicated by the fact that an important rival for ``transmundane identity'' has been proposed: David Lewis's \cite{Lewis1968,Lewis1986} counterpart theory. In this theory, the claim that an individual exists in more than one possible world replaces the claim that although each individual exists in only one world, it has counterparts in other worlds, where the counterpart relationship (based on similarity) does not satisfy the logic of identity. Therefore, much of the discussion in this area has concerned the comparative merits of transmundane identity and the theoretical explanations of the counterpart as interpretations, within a framework of possible worlds, of statements of what is possible and necessary for private individuals.
Usually, the identity of physical systems is taken for granted. More so, if we consider that, in standard quantum mechanics, the symmetrization postulate is an independent axiom, and one can discuss many particular physical phenomena (such as superpositions and entanglement) without taking it into account. However, the whole theory must be taken into account in discussions about ontology.

This analysis illustrates that we must consider quantum indistinguishability in the discussions about the interpretation of quantum mechanics. One either assumes, as in Bohmian mechanics, that quantum systems are individuals or one does not. If quantum systems are assumed to be non-individuals, then they cannot be identified in general. In particular, they cannot be identified when considered in different contexts. If we assume an ontology of non-individuals, the EPR argument is not entailed!

Identifying items (systems, properties, beables) among different (and mutually exclusive) possible worlds is problematic in general. For example, it would not be meaningful to identify the USA in our current world with a country in the alternative world of Phillip K. Dick's novel, where the Axis won WWII. One can compare both countries, the actual one, and the hypothetical one. However, it is meaningless to try to identify them. Identification among different contexts is even more problematic when systems are non-individuals, in the sense suggested by Schrodinger for quantum systems.

\subsection{The different steps that lead to the KS contradiction}

Let us go back again to the origin of the KS contradiction. One can identify the following steps:

\begin{itemize}
    \item Hidden variables are assumed assigning values to the different observables (represented by the self-adjoint operators in $\mathcal{A}(\mathcal{H})$).
    \item The values are assigned in so that they are \textit{context independent}. This means that the \textit{same} $\lambda$ defines the state of a single quantum system, and assigns the \textit{same} value $v_{\lambda}(A)$ to each observable $A$, independently of the context in which it is considered. In other words, the system is assumed to retain its identity among the different contexts. 
    \item Something similar happens with projection operators representing elementary properties: properties (with the same content) appearing in different contexts are identified, and their truth values are preserved.
    \item The above assumptions lead to the conclusion that there should exist a valuation from $\mathcal{L}(\mathcal{H})$ to the two-valued Boolean algebra, satisfying the functionality condition \ref{e:TruthValueResolutionIdentity}. However, no such valuation exists.
\end{itemize}

As we have seen in section \ref{s:Indistinguishability}, the assumption that quantum systems can be identified (and consistently re-identified) is problematic. This problem was also illustrated in the EPR example. As we can see, it is clear that an identification procedure is taken to derive the KS contradiction: the system is considered the same among the different contexts, and properties are concomitantly identified.

Let us now analyze the KS discussion under the metaphysical assumption that quantum systems are non-individuals. We can summarize the assumption leading to the KS contradiction as the following statement.
\begin{quote}
(KSH) \;  It is possible to assign context-independent and well-defined values to all
measurable properties of a \textit{single} quantum system.
\end{quote}
After what we have said above, it should be clear that we can avoid the KS contradiction by negating KSH in at least two ways.
\begin{description}
\item[(i)] Properties do not have well-defined context-independent truth values.
\item[(ii)] Properties or particles may be indistinguishable (due to the possibility that they are non-individuals).
\end{description}

The conclusion given by (i) is the most popular in the foundations of quantum mechanics. It is the usual way to avoid  the KS contradiction, as discussed in Section \ref{KSTheorem}. Similarly, it is argued that hidden variable theories, if they assign values to a quantum system's defining properties, must be contextual.  
However, we also have the second option. It may well be that option (ii) is true: if particles are truly indistinguishable entities, there is no way to identify the particles when considered in the different measurement contexts. Furthermore, this might have consequences for the measurement process: the intrinsic lack of particles' identity makes it meaningless to speak about properties as being properties of a specific particle. We think that these observations open the door to a novel interpretation of the KS result.

Under the non-individuality assumption, it seems odd to affirm that the properties defining a context $C$ correspond to the same particle than the ones defining a ``different" context $D$ prepared in the same way (an indistinguishable context, in the sense posed before). The act of choosing between measurements in contexts $C$ or $D$ corresponds to different (and usually incompatible) possible worlds. We cannot concede that we are talking about the same object (``the same reality", as in the EPR argument) underlying these alternatives: our non-individuality assumption implies that it is meaningless to assign a transworld identity to elementary particles. Notice that this argument needs not to be operational: it follows as a logical consequence of our ontological non-individuality assumption. There is no need to perform any actual experiments to realize that to affirm that we have the same particle in all contexts is a strong ontological assumption (dependent on the classical notion of identity).

\section{Conclusions}\label{s:Conclusions}

In this work, we have shown that one of the leading metaphysical assumptions behind the KS contradiction's derivation is that the systems are (covertly) considered \textit{individuals} who can be identified among the different measurement contexts. Consequently, it follows that properties retain their values because we are speaking about the \textit{same} physical system. Obviously, a similar analysis holds for most KS-like contradictions (a matter that we will tackle in future works).

However, as we have seen, if one takes earnestly quantum indistinguishability, it seems that one cannot so calmly assume that quantum systems are \textit{individuals}. At the very least, one must be explicit about this assumption and consider it when extracting ontological conclusions. Consequently, when we consider different measurement contexts, it is not clear that we are speaking about the same system. Of course, this is an interpretation-dependent issue since there are interpretations in which quantum systems are indeed individuals. We believe that considering explicitly an ontology in which quantum systems are truly indistinguishable entities --in the sense of Schr\"{o}dinger-- the previous discussions about interpretation --as we have shown for the case in the EPR argument-- provides an attractive perspective for ontological research. This perspective has the potential of being more attuned to the views of the physics community.


\begin{thebibliography}{999}

\bibitem{BvN}
Birkhoff, G.; von Neumann, J.  The Logic of Quantum Mechanics.
\emph{Ann. Math.}, {\bf 37}, 823--843 (1936).

\bibitem{KST}
Kochen, S.; Specker, E.P. The Problem of Hidden Variables in Quantum
Mechanics. \emph{J. Math. Mech.} \textbf{1967}, \textit{17}, 59--87.

\bibitem{Newton-Olimpia}
da Costa, N.; Lombardi, O.; Lastiri, M. A modal ontology of
properties for quantum mechanics. \emph{Synthese} \textbf{2013},
\textit{190}, 3671--3693.

\bibitem{Doring-KSVNA}
D\"{o}ring, A. Kochen-Specker Theorem for von Neumann Algebras.
\textit{Int. J. Theor. Phys.} \textbf{2005}, \textit{44}, 139--160.

\bibitem{Svozil-KS}
Svozil, K.; Tkadlec, J. Greechie diagrams, nonexistence of measures
in quantum logics, and Kochen-Specker-type constructions. \textit{J.
Math. Phys.} \textbf{1996}, \textit{37}, 5380--5401.

\bibitem{Smith-KS}
Smith, D.  Orthomodular Bell-Kochen-Specker Theorem. \textit{Int. J.
Theor. Phys.} \textbf{2004}, \textit{43}, 2023--2027.


\bibitem{AcacioContextuality-2015} de Barros, J.A.; Oas, G. \textit{Some Examples of Contextuality in Physics: Implications to Quantum Cognition---Contextuality from
Quantum Physics to Psychology}; World Scientific: Singapore, 2015.

\bibitem{Aerts}
Aerts, D.; Gabora, L.; Sozzo, S. Concepts and their dynamics: A
quantum-theoretic modeling of human thought. \textit{Top. Cogn.
Sci.} \textbf{2013}, \textit{5}, 737--772.

\bibitem{Khrennikiv-Ubiquitous}
Khrennikov, A.Y. \emph{Ubiquitous Quantum Structure From Psychology
to Finance}; Springer-Verlag, Berlin/Heidelberg, Germany, 2010.

\bibitem{Holik-Probabilities-Generalized} F. Holik, S. Fortin, G. Bosyk, A. Plastino (2017) On the Interpretation of Probabilities in Generalized Probabilistic Models. In: de Barros J., Coecke B., Pothos E. (eds) Quantum Interaction. QI 2016. Lecture Notes in Computer Science, vol 10106. Springer, Cham. 

\bibitem{Holik-2020-Non-individuals} F. Holik, J.P. Jorge and C. Massri.
``Indistinguishability right from the start in standard quantum
mechanics", arXiv preprint arXiv:2011.10903 (2020)

\bibitem{Schro-Non-individuals} E. Schrodinger,  What is an elementary particle? In Castellani, E.
(ed.) Interpreting Bodies: Classical and Quantum Objects in Modern
Physics. Princeton: Princeton Un. Press : 197-210, (1998).

\bibitem{French-Krause-2006} S. French and D. Krause. Identity in Physics: A Historical, Philosophical, and Formal Analysis. Oxford University Press, Oxford, 2006.

\bibitem{deBarros-Holik-Krause-2017}
de Barros, J.A.; Holik, F.; Krause, D. Contextuality and
Indistinguishability. \emph{Entropy} 2017, \textit{19}, 435,
doi:10.3390/e19090435.

\bibitem{DeBarros-Indistinguishability-2019} 
de Barros, J.A.; Holik, F. and Krause, D. Indistinguishability and the origins of contextuality in physics.
\textit{Philos. Trans. R. Soc. Math. Phys. Eng. Sci.}, Volume \textbf{377}, Number
\textbf{2157}, 20190150 (2019).

\bibitem{deBarros-IndistinguishabilityandNegative} 
de Barros, J.A. and Holik, F. Indistinguishability and Negative Probabilities. \textit{Entropy} 2020, \textbf{22}, \textbf{829}. https://doi.org/10.3390/e22080829

\bibitem{KolmogorovProbability}
Kolmogorov, A.N. \emph{Foundations of Probability Theory}; Julius
Springer: Berlin, Germany, 1933.


\bibitem{piron} Piron, C. {\it Foundations of Quantum Physics}; W.A.
Benjamin, New York, USA, 1976.

\bibitem{mikloredeilibro} R\'{e}dei, M.  \textit{Quantum Logic in Algebraic Approach}; Kluwer
Academic Publishers: Dordrecht, The Netherlands, 1998.

\bibitem{Redei-Summers2006}
R\'{e}dei, M.; Summers, S. Quantum probability theory. \textit{Stud.
Hist. Philos. Sci. Part Stud. Hist. Philos. Mod. Phys.}
\textbf{2007}, \textit{38}, 390--417.

\bibitem{Hamhalter}
Hamhalter, J. {Quantum Measure Theory}; In {\em Fundamental Theories
of Physics}; Kluwer Academic Publishers Group: Dordrecht, The
Netherlands,  2003; Volume~134.

\bibitem{Gleason}
Gleason, A.  Measures on the Closed Subspaces of a Hilbert Space.
\emph{J. Math. Mech.} \textbf{1957}, \textit{6}, 885--893.

\bibitem{Gleason-Dvurechenski-2009}
Buhagiar, D.; Chetcuti, E.; Dvure\v{c}enskij, A. On Gleason's
theorem without Gleason. \emph{Found. Phys.} \textbf{2009},
\textit{39}, 550--558.


\bibitem{vN}
von Neumann, J. {\it Mathematical Foundations of Quantum Mechanics},
12th ed.; Princeton University Press: Princeton, NJ, USA, 1996.




\bibitem{Isham} C.J. Isham and J. Butterfield. Topos Perspective on the Kochen-Specker Theorem: I. Quantum States as Generalized Valuations. Int. J. Theor. Phys. 1998, 37, 2669–2733.


\bibitem{Jorge-Holik-2020} J. P. Jorge and F. Holik. Non-Deterministic Semantics for Quantum States. \textit{Entropy}, \textbf{22}(2):156 (2020). 

\bibitem{Cabello-SimpleKS} A. Cabello, et al, \emph{Physics Letters A}, \textbf{212} (1996)
183-187.

\bibitem{Omar-2005} Y. Omar. Indistinguishable particles in quantum mechanics: an introduction, \textit{Contemporary Physics}, \textbf{46}:6, 437-448,  (2005). DOI: 10.1080/00107510500361274

\bibitem{HOM-1987} C. K. Hong, Z. Y. Ou and L. Mandel,\emph{Phys. Rev. Lett.} 59 (18): 2044-2046, (1987).

\bibitem{Omar-2002} N. Paunkovi\'{c}, Y. Omar, S. Bose and V. Vedral. Entanglement Concentration Using Quantum Statistics", \emph{Physical Review Letters}, Vol. \textbf{88}, Number \textbf{18} (2002)

\bibitem{LoFranco-PRL-2018} R. Lo~Franco and G. Compagno. ``Indistinguishability of elementary systems as a resource for quantum information
processing", \emph{Physical Review Letters}, \textbf{120}: 240403
(2018).

\bibitem{Castellini-2019} A. Castellini, R. Lo Franco, L. Lami, A. Winter, G. Adesso,
and G. Compagno. Indistinguishability-enabled coherence for quantum metrology.
\textit{Phys. Rev. A}, 100:012308, (2019).

\bibitem{EPR-paper} A. Einstein, B. Podolsky and N. Rosen. ``Can Quantum-Mechanical Description of Physical Reality be Considered omplete?". \textit{Physical Review}, \textbf{47} (10): 777-780 (1935).

\bibitem{Kripke} 
Kipke, S.Naming and Necessity, Oxford (1980): Basil Blackwell; this is the expanded monograph version of Kripke 1972.

\bibitem{Lewis1968}
Lewis, D., 1968, “Counterpart Theory and Quantified Modal Logic”, The Journal of Philosophy, 65: 113–126; reprinted in Loux 1979 and (with additional “Postscripts”) in Lewis 1983.

\bibitem{Lewis1986}
Lewis, D., 1986, On the Plurality of Worlds, Oxford: Basil Blackwell. (Sections 1–3 of Chapter 4 reprinted in Kim and Sosa 1999).








\end{thebibliography}
\end{document}